\documentclass[journal]{IEEEtran}

\usepackage{amsmath,amsfonts}
\usepackage{algorithmic}
\usepackage{algorithm2e}
\usepackage{array}
\usepackage{textcomp}
\usepackage{stfloats}
\usepackage{url}
\usepackage{verbatim}
\usepackage{graphicx}
\usepackage{enumitem}

\ifCLASSOPTIONcompsoc
 \usepackage[caption=false,font=normalsize,labelfont=sf,textfont=sf]{subfig}
\else
 \usepackage[caption=false,font=footnotesize]{subfig}
\fi

\usepackage{cite}
\usepackage{color}
\usepackage{bbold}
\usepackage{siunitx}
\usepackage{booktabs}
\usepackage{color,soul}
\usepackage{xcolor}

\begin{document}
\onecolumn{

\large
\noindent Please cite this paper as: \newline
M. Wojnar et al., ``IEEE 802.11bn Multi-AP Coordinated Spatial
Reuse with Hierarchical Multi-Armed Bandits,'' IEEE Communications Letters, doi: 10.1109/LCOMM.2024.3521079

\normalsize
\vspace{1cm}
\begin{verbatim}

@ARTICLE{10811972,
  author={Wojnar, Maksymilian and Ciezobka, Wojciech and Kosek-Szott, Katarzyna
  and Rusek, Krzysztof and Szott, Szymon and Nunez, David 
  and Bellalta, Boris},
  title={IEEE 802.11bn Multi-AP Coordinated Spatial Reuse 
  with Hierarchical Multi-Armed Bandits}, 
  journal={IEEE Communications Letters}, 
  year={2025},
  volume={29},
  number={3},
  pages={428-432},
  doi={10.1109/LCOMM.2024.3521079}}
\end{verbatim}

}
\twocolumn
\clearpage

\title{IEEE~802.11bn Multi-AP Coordinated Spatial Reuse with Hierarchical Multi-Armed Bandits}

%%% AUTHORS
% M&W
% K,K,S
% D,B

\author{Maksymilian Wojnar, Wojciech Ciezobka, Katarzyna Kosek-Szott, Krzysztof Rusek, Szymon Szott,\\David Nunez, Boris Bellalta
        % <-this % stops a space
\thanks{
M. Wojnar, W. Ciezobka, K. Kosek-Szott, K. Rusek, and S. Szott are with the AGH University of Krakow, Poland. Their work was funded by the National Science Centre, Poland (2020/39/I/ST7/01457). For the purpose of Open Access, the authors have applied a CC-BY public copyright license to any Author Accepted Manuscript (AAM) version arising from this submission. E-mails: name.surname@agh.edu.pl}
\thanks{
D. Nunez and B. Bellalta are with UPF Barcelona, Spain. Their work was partially funded by Wi-XR PID2021-123995NB-I00 (MCIU/AEI/FEDER,UE). E-mails: name.surname@upf.edu
}% <-this % stops a space
}



\maketitle

\begin{abstract} %75 to 100 words -> 88 is OK

Coordination among multiple access points (APs) is integral to IEEE 802.11bn (Wi-Fi 8) for managing contention in dense networks. This letter explores the benefits of Coordinated Spatial Reuse (C-SR) and proposes the use of reinforcement learning to optimize C-SR group selection. We develop a hierarchical multi-armed bandit (MAB) framework that efficiently selects APs for simultaneous transmissions across various network topologies, demonstrating reinforcement learning's promise in Wi-Fi settings. Among several MAB algorithms studied, we identify the upper confidence bound (UCB) as particularly effective, offering rapid convergence, adaptability to changes, and sustained performance.

\end{abstract}


% ---------------------------------
% ---------------------------------
% ---------------------------------
% ---------------------------------

\section{Introduction}

\IEEEPARstart{D}{ense} IEEE 802.11 network deployments, with multiple access points (APs), suffer from co-channel interference. 
Multi-AP coordination (MAPC) methods are a new class of solutions that involve intelligent sharing of resources between neighboring APs to improve their utilization, typically measured as an increase in overall network throughput \cite{giordano2023will}. 
MAPC will most likely be one of the main novel features included in Wi-Fi 8 (IEEE 802.11bn) \cite{asterjadhi2024motions}. The following variants are currently being developed~\cite{yu2024TGbn}: coordinated spatial reuse (C-SR), coordinated beamforming (C-BF), coordinated TDMA (C-TDMA), coordinated restricted target wake time (C-RTWT).
In this letter, we focus on the first of these.
% New protocols and frames will have to be designed for managing neighboring APs, including the exchange of channel and buffer state information between them. Communication between APs can be implemented either over-the-air, or using a wired back-haul network. In case a back-haul network is used, a server can be used to control the set of light-weight APs. 

\begin{figure}[!t]
\centering
\includegraphics[width=0.9\columnwidth]{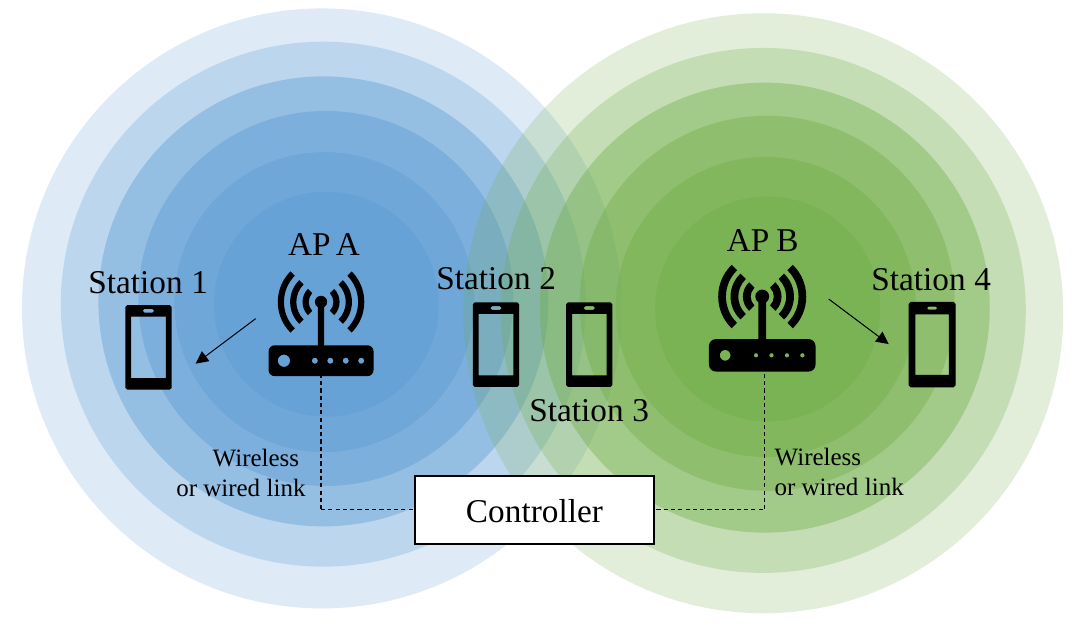}
% DISTRIBUTED CASE
%\caption{Example of C-SR: AP A wins channel access, becomes the sharing AP, selects the set of AP-station pairs (e.g., AP A to station 1 and AP B to station 4), informs AP B (a shared AP), and triggers the concurrent transmissions.}
\caption{Example of C-SR with a central controller: after AP~A wins channel access (becoming the sharing AP), the controller selects AP-station pairs (e.g., AP A to station 1 and AP B to station 4) for simultaneous transmission.}
\label{fig:mapc_csr}
\end{figure}

%% MAPC SR in more detail, with emphasis on the challenges: finding suitable groups +  scheduling
C-SR enables transmissions that are concurrent, on the same channel, and originate in neighboring overlapping networks (Fig.~\ref{fig:mapc_csr}).
Such transmissions can result in higher throughput and, potentially, lower delay by providing more transmission opportunities.
The coordination is done either by the AP that wins channel access (called the \textit{sharing AP}), which then triggers a subsequent frame exchange with the \textit{shared APs} \cite{nunez2022txop}, or through a central controller (which we assume here to study best-case performance gains).
These coordinated and concurrent transmissions make C-SR different from 802.11ax SR, in which Wi-Fi devices can transmit during ongoing transmissions from neighboring networks (if the received signal strength is low enough) \cite{verma2024survey}.
The signaling mechanisms required for \mbox{C-SR} are still being defined, so in the following we focus on the main outstanding challenge (determining which AP-station pairs are compatible, i.e., can be scheduled for concurrent transmissions) while assuming downlink transmissions.

The challenge of C-SR group selection, i.e., finding compatible AP-station pairs to perform C-SR transmissions, can be addressed by continuously collecting received signal strength (RSS) values from all network devices, exchanging them, and estimating compatible AP-station pairs~\cite{nunez2022txop}.
However, RSS values exhibit high variance (even in stationary scenarios)~\cite{kim2013smartphone}, and their exchange may cause significant overhead in larger networks~\cite{seok202coordinated}.
RSS measurements have been used as input for indoor localization and rate selection in Wi-Fi networks, but failed in both cases.
Therefore, our research is motivated by the following question: can we avoid the RSS exchange overhead by directly exploring which AP-station pairs are suitable for simultaneous transmissions? 
To answer this question, we turn to a reinforcement learning approach called multi-armed bandits (MABs).

MABs are lightweight, model-free algorithms that efficiently implement the trade-off between exploration (knowledge acquisition) and exploitation (knowledge application)~\cite{slivkins2019introduction}.
A MAB agent has limited resources: it can only make decisions a finite number of times, so it should discover the action with the highest reward as quickly as possible.
% An important constraint of using MABs for C-SR is the explosion of the number of options to explore, and therefore they are not suitable for solving complex tasks.
On the other hand, the main benefit of MABs is that they do not require \textit{a priori} knowledge of the environment and can learn online, which seems to suit the random nature of IEEE 802.11 deployments. 

In the literature, MABs are used to improve (uncoordinated) spatial reuse in overlapping IEEE 802.11 networks, where they find the best network parameter settings, both in centralized~\cite{bardou2021improving} and decentralized settings~\cite{wilhelmi2019potential,wilhelmi2019collaborative, kim2023improving, iturria2024cooperate}. 
Existing work focuses on channel selection, transmission power control, and adaptation of the sensitivity threshold for clear channel assessment. 
Commonly researched MAB algorithms are: $\epsilon$-greedy, Thompson sampling~(TS), upper confidence bound~(UCB), and Softmax~\cite{sutton_barto,slivkins2019introduction}.
% Furthermore, \cite{iturria2022cooperate} propose cooperation-based contextual MABs to improve spatial reuse. They show that with agent cooperation it is possible to improve SR in comparison to the non-cooperative scheme.
To reduce computational complexity, arising from a large search space and a continuous power set, we use hierarchical MABs, which have previously been successfully applied for resource allocation in satellite communication \cite{shen2023hierarchical}.

In this paper, we design a hierarchical MAB scheme for C-SR to select the best AP-station pairs effectively. 
To the best of our knowledge, this is the first paper to propose enhancing the upcoming C-SR feature with machine learning.
Our results show that MABs can quickly learn (i.e., in a few seconds) which pairs are the best for different network topologies. In detail, our contributions are as follows:
\begin{enumerate} 
    \item Definition of C-SR group selection as a MAB problem.
    \item Definition of a hierarchical MAB architecture -- a lightweight, model-free solution to this problem.
    \item Performance analysis of several MAB variants in \mbox{C-SR} scenarios that showcase their quick learning capabilities.
\end{enumerate}    
Furthermore, we provide a high-level C-SR simulator\footnote{\url{https://github.com/ml4wifi-devs/mapc-sim/tree/v0.1.6}}, released as open source for the research community.

\section{System Model}

To study %the applicability of MABs to 
the problem of C-SR group selection, we develop a Monte Carlo simulation model in Python that operates as follows.
First, from the positions of all nodes, we determine the RSS by calculating the path loss $P_{L}$ (in dB) from the TGax model for enterprise scenarios \cite{pathloss}:
\begin{equation}\label{pl_equation}
P_{L} = 40.05 + 20\log_{10}\left(\frac{\min(\delta,B_{p}) \cdot f_{c}}{2.4}\right) + \mathbb{1}_{\delta>B_p} P' + 7 W \text{,} \nonumber
\end{equation}
where $\delta$ is the distance between the transmitter and the receiver in meters (but no less than 1), $f_{c}$ is the carrier frequency in \SI{}{\giga\hertz}, $W$ is the number of walls traversed by the signal, and $P' = 35\log_{10}(\delta/B_{p})$  if $\delta$ is higher than the breaking point distance $B_{p}$ (otherwise, it is zero). We set $B_{p}$ to \SI{10}{\meter}.

Next, we account for interference from the set of concurrent transmissions $I$ and ambient noise $n$ (both in mW) to obtain the signal-to-interference and noise ratio (SINR). 
Additionally, to simulate SINR fluctuations, we add a normally distributed noise perturbation $\varepsilon$ with a standard deviation of \SI{2}{dB}. Finally, SINR is calculated as:
% \begin{equation}
%     \text{SINR} = T_X - \left( P_L + \log \left( \sum_{i \in I} \exp(i) + \exp(N) \right) \right) + \varepsilon,
% \end{equation}
\begin{equation}
    \text{SINR} = TX - \left( P_L + 10 \log \left( \sum I + n \right) \right) + \varepsilon, \nonumber
\end{equation}
where $TX$ is the transmission power in dBm. We then calculate the probability of successful transmissions for all AP-station pairs given the modulation and coding scheme (MCS) used for transmission. 
We approximate this probability using empirically obtained data, similarly to~\cite{ciezobka2023ftmrate}.
% We approximate this probability using curves obtained by fitting the cumulative distribution function of the normal distribution to empirically obtained data, %obtained from \mbox{ns-3}\footnote{\url{https://www.nsnam.org/}} simulations, 
% similarly to \cite{ciezobka2023ftmrate}. 
Our simulator iterates over consecutive transmission opportunities (TXOPs),
% , each governed by a sharing AP.
which have a duration $\tau$ set to the maximum value allowed by 802.11~\cite{kosek2019tuning}. Additionally, we implement the following simplified A-MPDU aggregation. We assume that the size of each A-MPDU sub-frame is fixed (Table~\ref{tab_simulation_settings}).
Additionally, for each TXOP we calculate the maximum number of aggregated MPDU sub-frames ($n_\text{A-MPDU}$) transmitted with a given data rate (dictated by the selected MCS) as:
\begin{equation}
 n_\text{A-MPDU}=\left\lceil\frac{\text{data rate} \times \tau}{\text{A-MPDU sub-frame size}}\right\rceil . \nonumber
\end{equation}
The resulting number of successfully received A-MPDU sub-frames is modeled using a binomial distribution with the calculated success probability and the number of trials equal to $n_\text{A-MPDU}$. 

As a performance metric, we measure the \textit{effective data rate}, 
calculated as the number of received bits divided by the duration of the TXOP and summed across all concurrent transmissions.
We assume that acknowledgments are transmitted in parallel using OFDMA to ensure that there are no collisions.
%Since we investigate the effective data rate, we do not address the problem of efficiently transmitting acknowledgments, although note that they could be sent through OFDMA to ensure a lack of collisions.
We also neglect the impact of additional signaling that MAPC will require as being outside of the scope of this article. %, where we focus on evaluating MAB performance.

% The underlying challenge of C-SR is selecting AP-station pairs.
% \section{Hierarchical MABs for C-SR}
\section{Proposed Algorithm}
\label{sec:mabs}

The C-SR group selection problem is the following: in each TXOP, the controller %sharing AP 
decides which AP-station pairs are suitable for simultaneous transmission. 
Formally, these are $(A_i \in \mathcal{A}, S_i^j \in \mathcal{S}_i)$ pairs, where $\mathcal{A}$ is the set of all $N=|\mathcal{A}|$ APs, and $\mathcal{S}_i$ is the set of stations associated with AP $A_i$.
% This decision is the main problem of C-SR and we propose using the hierarchical MAB scheme described in Section~\ref{sec:mabs}.

Our hierarchical MAB scheme (Algorithm~\ref{alg:mapc-mab}) is applied whenever any $A_k \in 
\mathcal{A}$ becomes a sharing AP.
This AP along with its associated station $S_k^l$, the recipient of the downlink transmission, create the first  pair $\mathcal{P}_0 \gets \{(A_k, S_k^l)\}$ (line~4), to which other pairs may be added as the algorithm is executed (lines~5 and~10). 
Since APs gain channel access according to legacy 802.11 rules (which ensure long-term fairness) and since we assume even traffic distribution, we select both $A_k$ and $S_k^l$ randomly and uniformly (lines~2--3).

\begin{figure}[t!]
    \centering
   \includegraphics[width=\columnwidth]{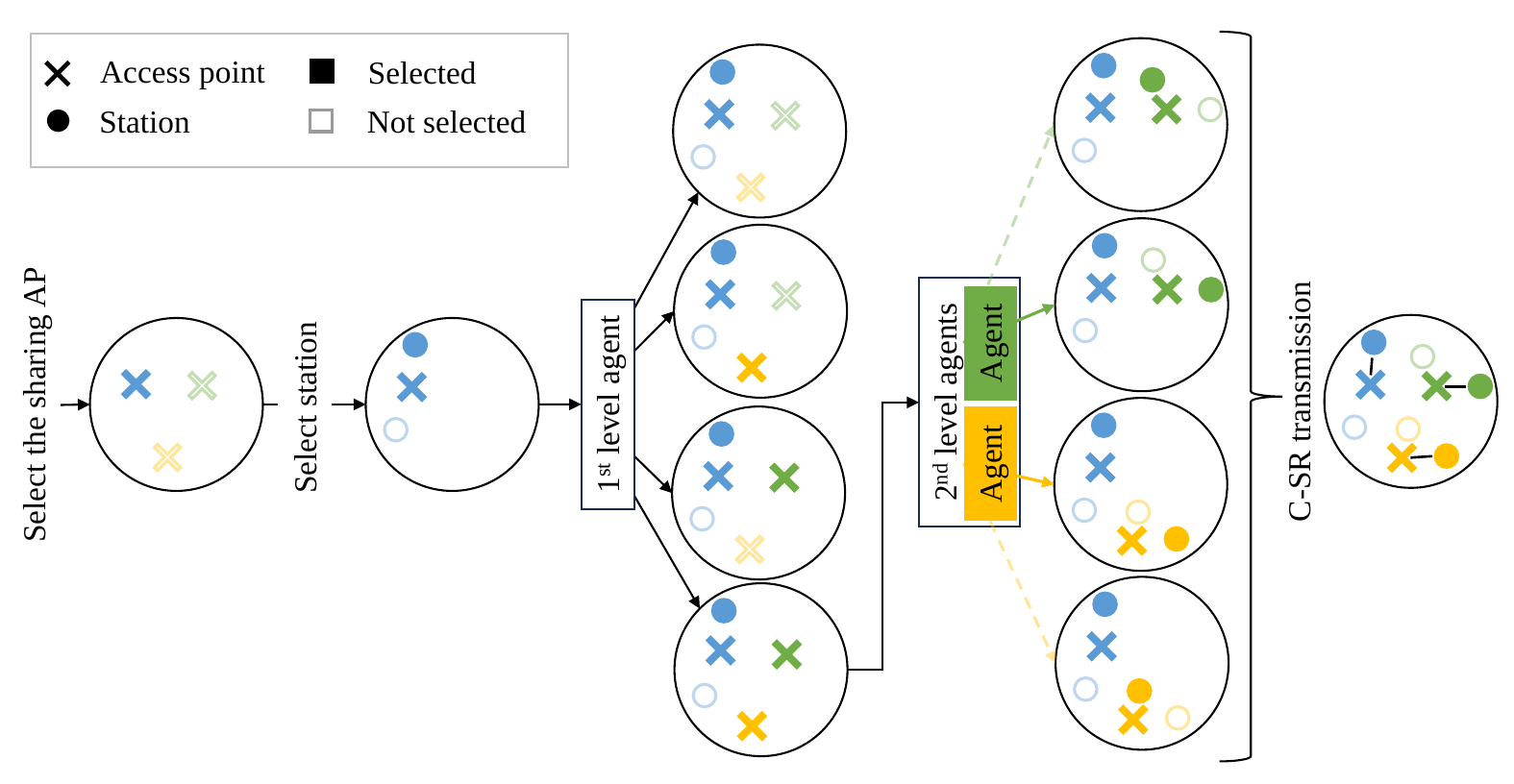}
    \caption{Example operation of the proposed hierarchical MAB scheme. Agents not taking part are omitted. The first level agent selects which other APs transmit. The two second level agents each select one of two recipient stations.}
    \label{fig:mab-scheme}
\end{figure}

The core of our hierarchical scheme is based on two levels of MAB agents.
The first level agents $\alpha^I$ select which other APs are to transmit in this TXOP (given $\mathcal{P}_0$). 
Then, the second level agents $\alpha^{II}$, given the set of transmitting APs, select  recipient stations for these APs (Fig.~\ref{fig:mab-scheme}).
Our hierarchical approach enables agents to leverage the structure of the C-SR problem and update their knowledge about the AP set (first level) after performing each action containing that set (second level).
This approach reduces the search space (in comparison to a non-hierarchical, flat MAB) and provides faster convergence.
Formally, the set $\alpha^I$ is 
% dedicated to station $S_i^j$, and
indexed by $(A_k, S_k^l)$
% The first-level agent assigned to station $S_k^l$ 
and selects a set $\mathcal{F}_k^l$ of shared APs (line~6). $\mathcal{F}_k^l$ together with $A_k$ form the set $\mathcal{F}$ of APs that transmit concurrently in the  TXOP (line~7).
Meanwhile, the set  $\alpha^{II}$ is indexed by $(A_i, \mathcal{F})$ and these agents select stations $S_i^j \in \mathcal{S}_i$ for each $A_i\in\mathcal{F}_k^l$ (line~9). % to establish AP-station pairs for coordinated downlink transmissions.
In total, there are $\sum_{i \in N}|\mathcal{S}_i|$ first level and $\sum_{\mathcal{F} \in 2^{\mathcal{A}}} |\mathcal{F}|$ second level agents.
We assume that all MAB agents (on both levels) are of the same type (e.g., $\varepsilon$-greedy) and have the same configuration (hyperparameter set $\theta$), fine-tuned as described in Section~\ref{sec:performance}.

\begin{figure}[t!]
    \centering
    \includegraphics[width=0.8\columnwidth]{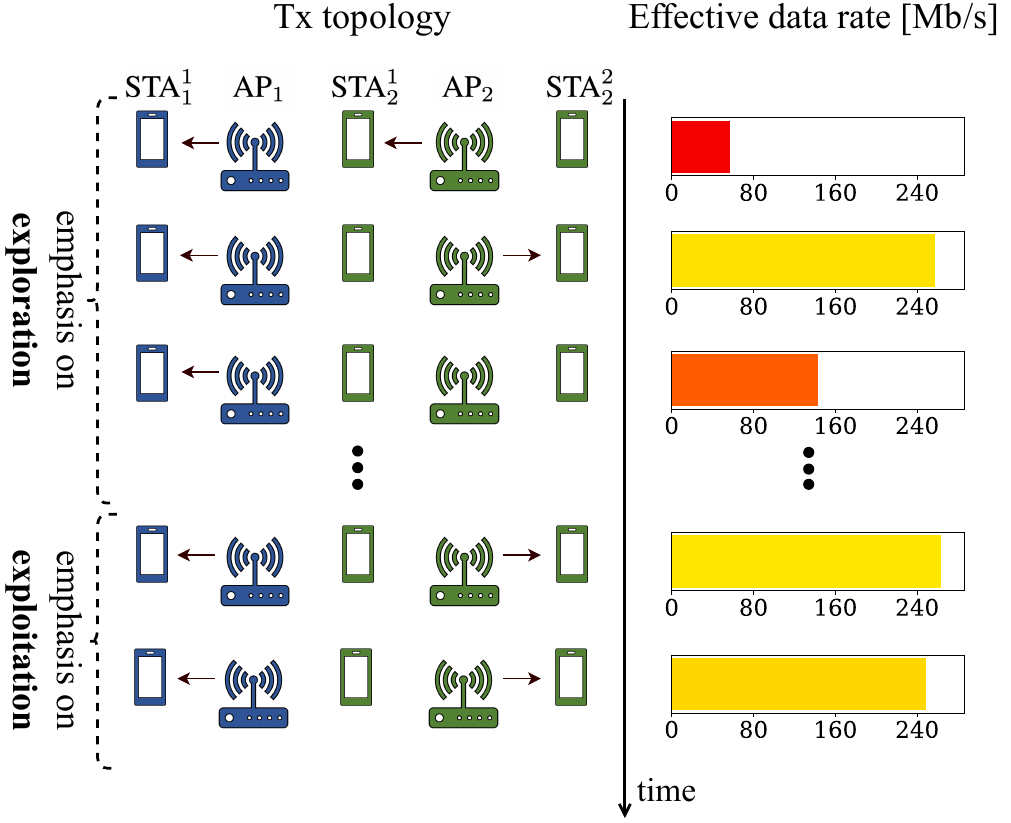}
    \caption{The agent's learning process when $\mathcal{P}_0$ is always $\{(A_1, S_1^1)\}$. Initially, the emphasis is on exploration to discover high-reward configurations. Over time, the agent gains confidence in which settings lead to high effective data rates and exploits them.
    }
    \label{fig:mab-diagram}
\end{figure}

Ultimately, we obtain a set $\mathcal{P}$ of AP-station pairs which perform simultaneous coordinated downlink transmissions (line~12). Then, we calculate the effective data rate (i.e., bits received across all parallel transmissions during the TXOP) which serves as the reward $r$ (line~13) used to update the agents (lines~14--17) and complete the learning process (Fig.~\ref{fig:mab-diagram}).

% \begin{figure}[t!]
%     \centering
%    \includegraphics[width=0.8\columnwidth]{Figures/mapc scheme(4).pdf}
%     \caption{Example operation of the proposed scheme.
%     }
%     \label{fig:mab-scheme}
% \end{figure}

\begin{algorithm}[t]
% \scriptsize
\caption{Hierarchical MAB scheme for C-SR.}
\label{alg:mapc-mab}
\KwIn{\\$\mathcal{A}=\{A_1,A_2,\ldots, A_N\}$ -- set of APs \\
$\mathcal{S}_i = \{S_i^1, S_i^2, \ldots, S_i^m\}$ -- stations associated with $A_i$ \\
$\mathcal{S} = \{\mathcal{S}_1, \mathcal{S}_2, \ldots, \mathcal{S}_N\}$ -- structure of APs and stations \\
$\theta$ -- agents' hyperparameters}
\textbf{Initialize:} \\
$\alpha^I \gets$ the first-level agents based on $\mathcal{A}, \mathcal{S}$ with $\theta$ \\ 
$\alpha^{II} \gets$ the second-level agents based on $\mathcal{A}, \mathcal{S}$ with $\theta$ \\ 
\textbf{Algorithm:} \\
\begin{algorithmic}[1]
\FOR{each TXOP}
% \STATE $\mathcal{P} \gets \emptyset$ 
\STATE Randomly and uniformly select $A_k \in \mathcal{A}$ % \gets$ 802.11 channel access, sharing AP
\STATE Randomly and uniformly select $S_k^l \in \mathcal{S}_k$ % \gets$ randomly selected, designated station
\STATE $\mathcal{P}_0 \gets \{(A_k, S_k^l)\}$ 
\STATE $\mathcal{P} \gets \mathcal{P}_0$
\STATE $\mathcal{F}_k^l \in 2^{\mathcal{A}\setminus\{A_k\}} \gets$ sample $\alpha^I \left[ \mathcal{P}_0 \right]$ %(shared APs)
%group sampled by the first level agent, given $(A_k, S_k^j)$
\STATE $\mathcal{F} \gets \mathcal{F}_k^l \cup \{A_k\}$
\FOR{$A_i \in \mathcal{F}_k^l$}
\STATE $S_i^j \in \mathcal{S}_i \gets$ sample $\alpha^{II} \left[ A_i, \mathcal{F} \right]$
%station sampled by the second level agent, given $(A_k, S_k^j)$ and $\mathcal{F}_k^j$
\STATE $\mathcal{P} \gets \mathcal{P} \cup \{(A_i, S_i^j)\}$
\ENDFOR
\STATE Perform simultaneous transmissions according to $\mathcal{P}$
\STATE $r \gets$ effective data rate in the TXOP
\FOR{$A_i \in \mathcal{F}_k^l$}
\STATE Update agent $\alpha^{II} \left[A_i, \mathcal{F} \right]$ with $r$
\ENDFOR
\STATE Update agent $\alpha^I \left[ \mathcal{P}_0 \right]$ with $r$
\ENDFOR
\end{algorithmic}
\label{alg1}
\end{algorithm}

% \section{Algorithm Deployment}

\begin{figure}[t!]
    \centering
    \includegraphics[width=0.85\columnwidth]{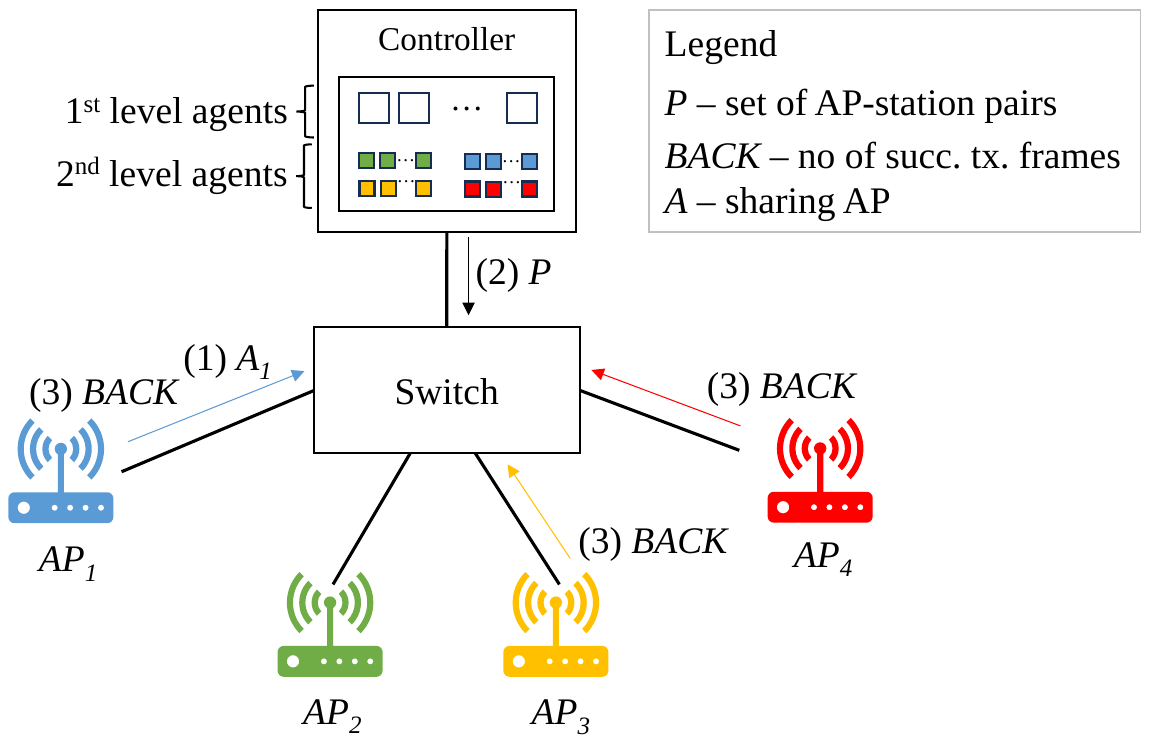}
    \caption{Example operation of proposed hierarchical MAB scheme with a central controller: (1) AP~1 notifies the controller that it is the sharing AP, (2) the controller provides the set $P$ of AP-station pairs to transmit in the next TXOP (which include APs~3 and~4), (3) after the TXOP all transmitting APs inform the controller how many frames were transmitted successfully.
    }
    \label{fig:centralized}
\end{figure}

In terms of deployment, our algorithm can operate both in a centralized and a decentralized regime. 
For brevity, we assume the former, i.e., all agents are deployed in a central network controller (Fig.~\ref{fig:centralized}).
This approach requires minimum signaling, which can be done over a wired connection, and limits all computational costs to the controller.
In the distributed case, the MAB agents have to be deployed to their corresponding APs. Also, additional communication is required
%Our algorithm can operate in a decentralized regime by deploying first and second-level MAB agents to corresponding APs. However, it needs communication 
between the APs to ensure that information about the result of the transmission is shared either over wired links or within MAPC control frames.
% Algorithm~\ref{alg:mapc-mab} provides a formal representation of the above steps.

% \begin{figure}[t!]
%     \centering
%     \includegraphics[width=0.9\columnwidth]{Figures/MAPC MAB diagrams-Centralized.drawio.png}
%     \caption{Centralized operation of hierarchical MAB scheme.
%     }
%     \label{fig:centralized}
% \end{figure}

\section{Performance Evaluation}
\label{sec:performance}

\begin{table}
\footnotesize
\caption{Simulation parameter settings}
\label{tab_simulation_settings}
\centering
\begin{tabular}{@{}ll@{}}
\toprule
\textbf{Parameter}         & \textbf{Value}            \\ \midrule
Band                       & 5   GHz                   \\
PHY                        & IEEE   802.11ax           \\
Channel   width            & 20   MHz                  \\
Spatial   streams          & 1, SISO                   \\
MCS                        & 11                        \\
Guard   interval           & 800   ns                  \\
Frame   aggregation        & Enabled (A-MPDU)                 \\
TXOP duration $\tau$              & \SI{5.484}{\milli\second} \\
Loss   model               & TGax w/ additive white Gaussian noise \\
Transmission power         & \SI{16.0206}{dBm}         \\
Noise floor                & \SI{-93.97}{dBm}          \\
Per-station   traffic load & Downlink, full buffer     \\
A-MPDU sub-frame   size        & 1500   B                  \\ \bottomrule
\end{tabular}
\end{table}

\begin{figure} 
    \centering
  \subfloat[Topology\label{fig:scenario_loc}]{%
       \includegraphics[width=0.4\linewidth]{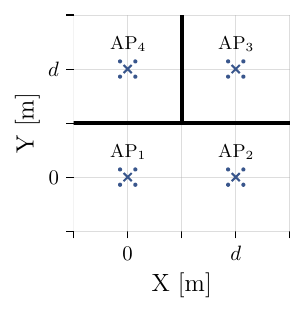}}
    \hfill
  \subfloat[Transmission strategies\label{fig:alignment}]{%
        \includegraphics[width=0.55\linewidth]{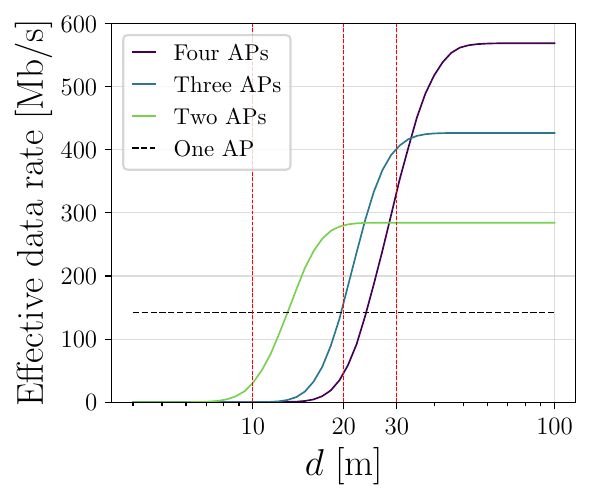}}
  \caption{Test scenario. In (a), crosses denote APs, dots -- stations, thick lines -- walls,  APs are placed on the corners of a $d$-sided square while stations are placed $\SI{2}{\meter}$ from their APs in  ordinal directions. In (b), the effective data rates of the coordinated transmissions (number of concurrently transmitting APs) depend on the square side $d$, assuming a fixed MCS of 11. Three operational points studied in further simulations are denoted by vertical lines.}
  \label{fig1} 
\end{figure}

To conduct a performance analysis, each agent in our hierarchical MAB scheme has to follow a certain logic (i.e., MAB algorithm) to make the necessary decisions.
We select four commonly researched MAB algorithms: $\epsilon$-greedy, TS (with a normal distribution), UCB, and Softmax
and use their implementations  from~\cite{wojnar2023reinforced}.
Since all MAB agents (on both levels) use the same algorithm, we study the performance of these algorithms separately.
Additionally, we use two baselines for comparison:
single transmissions (i.e., in each TXOP there is exactly one concurrent transmission) and a flat MAB algorithm. The latter is a non-hierarchical approach in which flat agents use one of the four above-mentioned algorithms;  we present only the best performing algorithm for brevity.
Furthermore, in the performance analysis, we focus on downlink transmissions, with full per-station buffers, fixed rate selection, and transmission power.
All relevant parameter settings are given in Table~\ref{tab_simulation_settings}.

Before evaluating agent performance in the test scenario (Fig.~\ref{fig1}), we first need to (separately) optimize the hyperparameters of all MAB algorithms. 
For this optimization, we use Optuna\footnote{\url{https://optuna.readthedocs.io}} in randomly generated scenarios with varying configurations (2--5 randomly placed APs on a square of size $\SI{75}{\meter} \times \SI{75}{\meter}$, each with 3--5 associated stations placed randomly around the APs using independent Gaussian distributions with a standard deviation between \SI{4}{\meter} and \SI{8}{\meter}. 
Furthermore, to adapt agents for non-stationary environments, the positions of APs and stations are altered several times during each run. This training, conducted in various scenarios, aims to capture a broad spectrum of potential real-world conditions.
The resulting hyperparameter values are included in our repository\footnote{\url{https://github.com/ml4wifi-devs/mapc-mab/tree/v1.0.0}}.

As our test scenario, we use the enterprise topology of Fig.~\ref{fig:scenario_loc}, 
% where four APs are placed at the corners of a $d$-sided square.
where
the walls are placed asymmetrically, allowing for the analysis of one, two, three, or four coordinated transmissions.
We first evaluate a best-case scenario, where APs transmit only to the outermost stations (e.g., AP4 to its top-left station), which receive the least amount of interference from other transmitting APs. %For example, for the two transmitting APs case we use a fixed selection of APs which are furthest apart (on the diagonal) and also select their stations which are furthest apart. 
%e.g., for two transmitting APs we use a fixed selection of APs which are furthest apart (on the diagonal) and also select their stations which are furthest apart.
% \st{This approach ensures an upper bound at the cost of fairness.}
As $d$ varies, so does the best transmission strategy (Fig.~\ref{fig:alignment}).
The baseline is the case where only one AP transmits,
which obviously performs the same regardless of $d$, and is the best strategy when all APs are close together.
As the APs are farther apart and the interference from ongoing transmissions decreases, more APs can successfully transmit at the same time.
These results denote an `upper bound', since we consider only a single set of AP-station pairs in each TXOP.

From this preliminary evaluation, we select three points of interest ($10$, $20$, and \SI{30}{\meter}) allowing one, two, or three simultaneous transmissions, respectively, to see (in the following) whether MABs can learn the best coordinated strategy.
To add dynamics to the system, the stations begin at \SI{2}{\meter}  from their APs, but halfway through the simulation they relocate to either \SI{3}{\meter} or \SI{4}{\meter} from the AP in the second and third cases, respectively.

\setbox0=\hbox{%
  \includegraphics[height=0.155\textwidth]{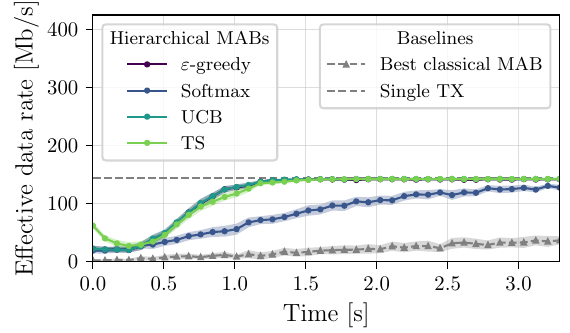}%

}
\setbox1=\hbox{%
  \includegraphics[height=0.155\textwidth]{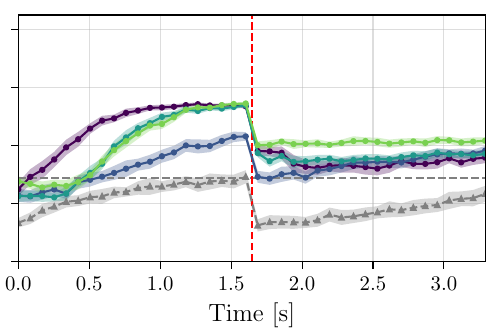}%
}
\setbox2=\hbox{%
  \includegraphics[height=0.155\textwidth]{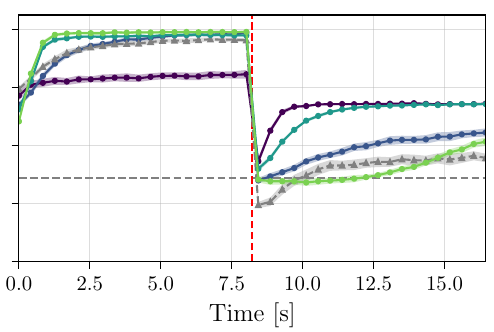}%
}
\newdimen\fght
\fght=\ht0
\begin{figure*}[!ht]
%\captionsetup[subfigure]{labelformat=empty}
\centerline{%
\hbox{%\hskip1cm%
  \vbox{\vskip0pt%
    \hbox{%
      \subfloat[$d = 10$ m \label{fig:data_rates-1}]{%
        \includegraphics[height=0.155\textwidth]{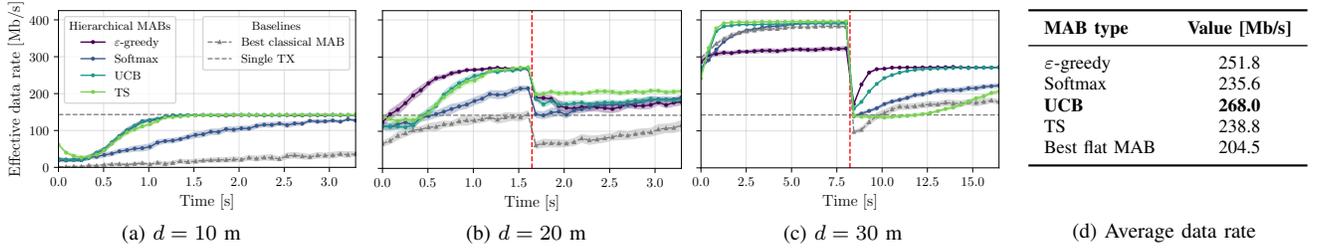}
      }
      \subfloat[$d = 20$ m \label{fig:data_rates-2}]{%
        \box1
      }
      \subfloat[$d = 30$ m \label{fig:data_rates-3}]{%
        \box2
      }
      \vspace{-0.5cm}
    }%
  }\hskip2mm%
  \vbox to \fght{\vskip0pt\vfil
  \vspace{-0.2cm}
    \hbox{\subfloat[Average data rate \label{tab:subkey}]{%
    \scriptsize
    \begin{tabular}[h]{lc}
         \toprule
         \textbf{MAB type}         & \textbf{Value [Mb/s]}  \\
         \midrule
         $\varepsilon$-greedy & 251.8 \\ % 5800.1          \\
         Softmax              & 235.6 \\ %5425.6          \\
         \textbf{UCB}                  & \textbf{268.0} \\ % \textbf{6172.9}          \\
         TS                   & 238.8 \\ %5501.0          \\
         Best flat MAB        & 204.5 \\ %4710.3          \\
         \bottomrule
         \vspace{0.3cm}
    \end{tabular}}%
    }\vfil%
  }%
}
}
\vspace{0.3cm}
\caption{Effective data rate achieved by agents in our hierarchical MAB scheme (Algorithm~\ref{alg:mapc-mab}), for various distances between APs, compared to a single coordinated transmission and to non-hierarchical (flat) MABs. %The figures present the results for node settings at different distances. 
In (c), we present a longer simulation run. The bands represent 99\% confidence intervals from 40 runs, while the red line indicates when stations relocate in (b) and (c). The average data rate across all tested cased (a)--(c) is given in (d).}
\label{fig:scenario_results}
\end{figure*}

% \begin{figure*}[!t]
% \centering
% \includegraphics[width=\textwidth]{Figures/Results/data-rates.pdf}
% \vspace{-0.9cm}
% \caption{Effective data rate achieved by agents in our hierarchical MAB scheme (Algorithm~\ref{alg:mapc-mab}), for various distances between APs, compared to a single coordinated transmission and to non-hierarchical (flat) MABs. %The figures present the results for node settings at different distances. 
% In (c), we present a longer simulation run. The bands represent 99\% confidence intervals from 40 simulation runs, while the red line indicates the point of station relocation in the second and third cases.}
% \label{fig:scenario_results}
% \end{figure*}

The 802.11 channel access rules give all APs equal transmission opportunities. So we first check if MABs can properly select the `single transmitting AP' configuration, which is the best coordination option when $d=\SI{10}{\meter}$.
Initially, MABs explore multi-AP options but converge on the single AP solution within $1$--$\SI{1.5}{\second}$ (Fig.~\ref{fig:scenario_results}a).
At \SI{20}{\meter}, simultaneous transmissions (regardless of destination) by two APs placed on the diagonal of the $d$-sided square can improve the effective data rate, and MABs quickly converge to this option (Fig.~\ref{fig:scenario_results}b), especially $\varepsilon$-greedy. Halfway through the simulation, the stations move away from the APs, which lowers the probability of a successful transmission. Nevertheless, this topology still allows for achieving a higher throughput compared to performing only single uncoordinated transmissions. %, by sticking to the same actions as before. 
MABs quickly adapt to this non-trivial configuration, and TS performs particularly well in this case.
Similarly, a further improvement resulting from three concurrent transmissions is achieved in the first phase of the last scenario (Fig.~\ref{fig:scenario_results}c).
These results come from a longer run to show that, over time, MABs can slightly improve the achieved rates. Note that in this case, the mobility of the stations significantly disrupts the reward distribution to which MABs have adapted. It causes a temporary drop in performance, but agents quickly find the new optimal configuration. UCB performs particularly well in this scenario.

To prove the advantage of our hierarchical approach, we implement and measure the performance of a flat, non\nobreakdash-hierarchical MAB, where the set of pairs that transmit in parallel to the given first pair $\mathcal{P}_0=\{(A_k, S_k^l)\}$ is determined in a single step. The same algorithms and hyperparameter optimization methods were used to ensure fairness. For clarity, only the best performing algorithms were selected for comparison. The performance and convergence of flat MABs (dashed lines with triangles in Fig.~\ref{fig:scenario_results}) are inferior to the hierarchical approach because the agent must reject each action containing pair $(A_i, S^j_i)$ separately, even if the AP $A_i$ is topologically incompatible with $A_k$ in terms of parallel transmission.

When comparing the performance of the MAB algorithms in our hierarchical approach, we first observe that $\varepsilon$--greedy has a short convergence in the first stage due to intensive exploration at the beginning of its operation. However, it does not always find the optimal configuration on the first attempt.
%Comparing the performance of the MAB algorithms, we first observe that $\varepsilon$--greedy has a short convergence due to intensive exploration in the first phase of the operation; this feature is referred to in the literature as ``optimistic start''. It also consistently selects the best-valued actions due to the low exploration factor.
% Comparing the performance of the MAB algorithms, we first observe that $\varepsilon$--greedy has a short convergence due to intensive exploration at the beginning of the operation, which rapidly decreases with the time; After that, it  consistently selects the best-valued actions learnt.
TS can easily discover the best solution and slightly outperform other approaches in the long, static run. Since TS strengthens its certainty over time, its adaptation at the late stage can be quite slow (cf. Fig.~\ref{fig:scenario_results}c).
UCB, as an asymptotically optimal algorithm, strikes a balance between exploration and exploitation, is the best general-purpose choice in terms of our tests achieving the best average performance over all tested cases.
Softmax behaves slightly worse in some scenarios, both in terms of the convergence time and effective data rate achieved. 
Since Softmax demonstrates an increased sensitivity to parameter adjustments, it can achieve either rapid convergence or maintain effectiveness over an extended period, and during optimization, we tried to find a balance between swift convergence and peak performance.
Furthermore, we evaluated fairness in channel access and found (though omit the numerical results for brevity) that with our C-SR solution all stations have more transmission opportunities than in a theoretical single transmission scheme with ideal fairness.
Overall, we conclude that MABs, after appropriate hyperparameter configuration, can typically rapidly converge on the best C-SR configurations in a given topology. In real testbeds, such hyperparameter configuration can be achieved within the digital twin paradigm.
Finally, we note that UCB is straightforward to implement and excels in performance: it quickly converges, adapts to rapid changes, and exhibits long-term effectiveness. 
This statement is confirmed by the average data rates achieved by each type of agent across the three presented cases (Fig.~\ref{fig:scenario_results}d).
This average is only one of many metrics that can be used to compare agents, but it considers both convergence time and steady-state performance.

% To confirm this statement, in Fig.~\ref{fig:scenario_results}(d), we present the total data transfer achieved by each
% %aggregated performance as a per-MAB sum of integrals 
% MAB-based C-SR calculated as a sum of the data transfer achieved for the cases in Fig.~\ref{fig:scenario_results}(a)-(c).
% %Thus, Fig.~\ref{fig:scenario_results}(d) shows the total amount of data (in Mb) successfully transmitted by APs when controlled by different MAB agents.

% % ALGOS

% \begin{table*}[t]
%     \centering
% \caption{Comparison of MAB algorithms}
% \begin{tabular}{ r|c |c |c| c}
% & \textbf{$\epsilon$-greedy} & \textbf{Softmax} & \textbf{UCB}& \textbf{NormalTS} \\
% \midrule
% %$\mathrm{init}$ & $S_0=(Q=1,N=1)$ & & &  \\  
% %$\mathrm{update}(S_t,A_t,R_t)$ & $Q_{t + 1}(a) = Q_t + \frac{1}{t} \lbrack R_t - Q_t \rbrack, N_{t+1}(a)=N_{t}(a)+1$& & &  \\ 
% %$\mathrm{sample}(S_t)$ & $\sim (1-\epsilon) \mathcal{D}(\arg\max Q) +\epsilon \mathcal{U}(Q)$ & & &  \\  
% agent state  & expected reward & preference (negative energy)& total reward& normal conjugate prior parameters \\
% $\mathrm{update}(S_t,A_t,R_t)$ & online average & stochastic gradient ascent & reward accumulator & conjugacy \\
% $\mathrm{sample}(S_t)$ & mixture of $\arg\max$ and uniform& Boltzman distribution & deterministic & $\arg\max$ on sample from posterior
% \end{tabular}
%     \label{tab:mab_algorithms}
% \end{table*}

\section{Conclusion}

This paper addresses the challenge of enabling C-SR in dense IEEE 802.11 network deployments, focusing on determining suitable AP-station pairs for concurrent transmissions. To solve this challenge, the paper proposes a novel hierarchical MAB architecture, an approach based on reinforcement learning in a new 802.11 setting.
The results highlight the ability of the proposed architecture to quickly learn the most suitable AP-station pairs for different network topologies within a few seconds. 

In general, this research contributes to the advancement of MAPC methods, particularly in the context of IEEE 802.11bn, and underscores the potential of machine learning to optimize resource utilization in dense wireless networks. As future work, we will %focus adapting transmission power and data rate selection.  as well as methods to reduce the number of `arms', e.g., using collected RSSI information to enhance agent's decisions. We will also 
consider variable transmission power settings, non-saturated traffic, data rate selection, and the impact of MAPC overheads. Finally, we will conduct an experimental validation to confirm the simulation results in reality.

\bibliographystyle{IEEEtran}
\bibliography{bibliography}

\begin{thebibliography}{10}
\providecommand{\url}[1]{#1}
\csname url@samestyle\endcsname
\providecommand{\newblock}{\relax}
\providecommand{\bibinfo}[2]{#2}
\providecommand{\BIBentrySTDinterwordspacing}{\spaceskip=0pt\relax}
\providecommand{\BIBentryALTinterwordstretchfactor}{4}
\providecommand{\BIBentryALTinterwordspacing}{\spaceskip=\fontdimen2\font plus
\BIBentryALTinterwordstretchfactor\fontdimen3\font minus \fontdimen4\font\relax}
\providecommand{\BIBforeignlanguage}[2]{{%
\expandafter\ifx\csname l@#1\endcsname\relax
\typeout{** WARNING: IEEEtran.bst: No hyphenation pattern has been}%
\typeout{** loaded for the language `#1'. Using the pattern for}%
\typeout{** the default language instead.}%
\else
\language=\csname l@#1\endcsname
\fi
#2}}
\providecommand{\BIBdecl}{\relax}
\BIBdecl

\bibitem{giordano2023will}
L.~Galati-Giordano, G.~Geraci, M.~Carrascosa, and B.~Bellalta, ``{What will Wi-Fi 8 Be? A Primer on IEEE 802.11bn Ultra High Reliability},'' \emph{IEEE Communications Magazine}, vol.~62, no.~8, pp. 126--132, 2024.

\bibitem{asterjadhi2024motions}
A.~Asterjadhi, ``{TGbn Motions List -- Part 1},'' Jul. 2024, doc.: 802.11-24/171r15.

\bibitem{yu2024TGbn}
R.~J. Yu, ``{Specification Framework for TGbn},'' Nov. 2024, doc.: IEEE 802.11-24/0209r6.

\bibitem{nunez2022txop}
D.~Nunez, F.~Wilhelmi, S.~Avallone, M.~Smith, and B.~Bellalta, ``{TXOP sharing with coordinated spatial reuse in multi-AP cooperative IEEE 802.11be WLANs},'' in \emph{Proc. of IEEE CCNC}, 2022.

\bibitem{verma2024survey}
S.~Verma \emph{et~al.}, ``{A Survey on Multi-AP Coordination Approaches Over Emerging WLANs: Future Directions and Open Challenges},'' \emph{IEEE Communications Surveys \& Tutorials}, vol.~26, no.~2, pp. 858--889, 2024.

\bibitem{kim2013smartphone}
Y.~Kim, H.~Shin, Y.~Chon, and H.~Cha, ``{Smartphone-based Wi-Fi tracking system exploiting the RSS peak to overcome the RSS variance problem},'' \emph{Pervasive and Mobile Computing}, vol.~9, no.~3, 2013.

\bibitem{seok202coordinated}
Y.~Seok \emph{et~al.}, ``{Coordinated Spatial Reuse (C-SR) Protocol},'' May 2020, doc.: IEEE 802.11-20/0576r1.

\bibitem{slivkins2019introduction}
A.~Slivkins, ``Introduction to multi-armed bandits,'' \emph{Foundations and Trends{\textregistered} in Machine Learning}, vol.~12, no. 1-2, pp. 1--286, 2019.

\bibitem{bardou2021improving}
A.~Bardou \emph{et~al.}, ``{Improving the spatial reuse in IEEE 802.11ax WLANs: A multi-armed bandit approach},'' in \emph{ACM MSWiM}, 2021.

\bibitem{wilhelmi2019potential}
F.~Wilhelmi \emph{et~al.}, ``{Potential and pitfalls of multi-armed bandits for decentralized spatial reuse in WLANs},'' \emph{Journal of Network and Computer Applications}, vol. 127, pp. 26--42, 2019.

\bibitem{wilhelmi2019collaborative}
------, ``Collaborative spatial reuse in wireless networks via selfish multi-armed bandits,'' \emph{Ad Hoc Networks}, vol.~88, 2019.

\bibitem{kim2023improving}
H.~Kim, G.~Na, H.~Im, and J.~So, ``{Improving Spatial Reuse of Wireless LANs using Contextual Bandits},'' \emph{IEEE Transactions on Wireless Communications}, vol.~23, no.~7, pp. 6735--6749, 2024.

\bibitem{iturria2024cooperate}
P.~E. Iturria-Rivera, M.~Chenier, B.~Herscovici, B.~Kantarci, and M.~Erol-Kantarci, ``Cooperate or not cooperate: Transfer learning with multi-armed bandit for spatial reuse in wi-fi,'' \emph{IEEE Transactions on Machine Learning in Communications and Networking}, 2024.

\bibitem{sutton_barto}
R.~S. Sutton and A.~G. Barto, \emph{{Reinforcement Learning: An Introduction}}.\hskip 1em plus 0.5em minus 0.4em\relax Cambridge, MA, USA: A Bradford Book, 2018.

\bibitem{shen2023hierarchical}
L.-H. Shen \emph{et~al.}, ``{Hierarchical Multi-Agent Multi-Armed Bandit for Resource Allocation in Multi-LEO Satellite Constellation Networks},'' in \emph{IEEE VTC2023-Spring}, 2023.

\bibitem{pathloss}
S.~Merlin \emph{et~al.}, ``{TGax Simulation Scenarios},'' Nov. 2015, doc.: IEEE 802.11-14/0980r16.

\bibitem{ciezobka2023ftmrate}
W.~Ciezobka, M.~Wojnar, K.~Kosek-Szott, S.~Szott, and K.~Rusek, ``{FTMRate: Collision-Immune Distance-based Data Rate Selection for IEEE 802.11 Networks},'' in \emph{Proc. of IEEE WoWMoM}, 2023.

\bibitem{kosek2019tuning}
K.~Kosek-Szott and N.~Rapacz, ``{Tuning Wi-Fi traffic differentiation by combining frame aggregation with TXOP limits},'' \emph{IEEE Communications Letters}, vol.~24, no.~3, pp. 700--703, 2019.

\bibitem{wojnar2023reinforced}
M.~Wojnar, S.~Szott, K.~Rusek, and W.~Ciezobka, ``{R}einforced-lib: {R}apid prototyping of reinforcement learning solutions,'' \emph{SoftwareX}, vol.~26, p. 101706, 2024.

\end{thebibliography}

\end{document}